\documentclass[12pt]{iopart}
 \pdfoutput=1 
\usepackage{graphicx}
\usepackage{amssymb}
\usepackage{color}
\usepackage{cite}   
\DeclareGraphicsExtensions{.pdf,.png,.jpg} 

\begin{document}
\newcommand{\wq}[1]{\textcolor{blue}{#1}}

\title{Universal behavior for single-file diffusion on a 
disordered fractal}

\author{L.~Padilla, H.~O.~M\'artin, J.~L.~Iguain}
\address{Instituto de Investigaciones F\'{\i}sicas de Mar del Plata (IFIMAR) 
and\\
Departamento de F\'{\i}sica FCEyN,
Universidad Nacional de Mar del Plata,\\
De\'an Funes 3350, 7600 Mar del Plata, Argentina}

\pacs{05.40.-a, 05-40.Fb, 66.30.-h}

\begin{abstract}
We study single-file diffusion on a one-dimensional lattice with a random
fractal distribution of hopping rates. For finite lattices, this  problem shows 
three clearly different regimes, namely, nearly independent particles, 
highly interacting particles, and saturation. 
The mean-square displacement of a tagged particle as a function of time 
follows a power law in each regime.
The first crossover time $t_s$, between the first and the second regime,
depends on the particle density. The other crossover time $t_l$, 
between the second and the third regime, depends on the lattice length.
We find analytic expressions for these dependencies and show how 
the general behavior can be characterized by an universal form. We
also show that the mean-square displacement of the center of mass
presents two regimes; anomalous diffusion for times shorter 
than $t_l$, and normal diffusion for times longer than $t_l$.
\end{abstract}
\maketitle

\section{Introduction}\label{sec:intro}

The subdiffusive behavior of a single random walk (RW) in a fractal
medium has been explained many year ago, as the effect of
obstacles of all sizes, which slow down the dynamics at every time 
scale~\cite{havlin1987,rammal1983,alexander1982,bouchaud1990,ben2000diffusion}.
More recently, considerable effort has been dedicated to study 
single-particle diffusion in finitely ramified  self-similar 
and self-affine media. It has been repeatedly shown that, in this cases,
the time behavior of the particle mean-square displacement (MSD) follows
a subdiffusive power-law modulated by logarithmic-periodic 
oscillations~\cite{Grabner1997, acedo2000, bab2008EPL, bab2008JCP, 
maltz2008, weber2010, Padilla2009, lorena2010, Padilla2011, havlin2000, bernhard2004, 
Padilla2012,refId0}. 

Subdiffusion can also result from the interaction between particles, even in
a substrate without  any hierarchical structure. This is, for instance, the
situation of a set of particles diffusing on a one-dimensional medium, in such
a way that the interactions prevent for particles to jump over one another.
The sequence of particles is then preserved, and the phenomenon is commonly 
called single-file diffusion~\cite{Harris1965, Percus1974, Richards1977,
Beijeren1983,  Lizana2009, Centres2010, Suarez2013 }. In the case of 
single-file diffusion on
an infinite homogeneous line, a representative tagged particle always
starts diffusing normally, provided that interactions are of short range. 
However, after a finite time, the dynamics become
subdiffusive, and the tagged-particle MSD behaves as $\sim t^{1/2}$ as a
function of time $t$, because of the correlations between neighbor 
particles~\cite{Suarez2013}. The tagged-particle MSD recovers the normal 
behavior ($\sim t$) at long times if single-file diffusion occurs 
on a homogeneous  segment of finite size with periodic  boundary conditions. This second crossover is closely related to existence
of a growing correlation length, which becomes of the order of
substrate size at the crossover time.
This three-regime behavior 
can be expressed in a universal form, where the crossover times depend on
the system size, average particle concentration, and interaction 
properties~\cite{Centres2010}. 

Single-file diffusion on a fractal has been analyzed in Ref.~\cite{Suarez2014},
for a set of hard-core interacting particles moving on a one-dimensional
lattice with a self-similar distribution of hopping-rates~\cite{Padilla2009}.
In this problem, after a finite time, the tagged-particle MSD shows  global
subdiffusive behavior ($\sim t^\nu$, with $\nu<1/2$) modulated by 
logarithmic-periodic oscillations. The attempt to find an universal form fails
because the modulations do not collapse on a single curve when scaling with
respect to a single variable.

In this article, we address the problem of single-file diffusion for
a set of hard-core interacting particles diffusing on a one-dimensional 
disordered fractal, which can be obtained by randomly shuffling the hopping
rates of a self-similar lattice. As for the case of non-interacting
particles~\cite{Padilla2011}, we show that the disorder washes the 
oscillations out.  As a generalization of the homogeneous case, the 
tagged-particle MSD presents  three regimes obeying a universal
scaling form, which reflects the interplay between fractal 
properties of the substrate and hard-core interactions.
We also study the center-of-mass MSD, which shows two regimes, subdiffusion 
at short times, and normal diffusion at long times; and is governed by another 
universal scaling law.

The paper is organized as follows. In section~\ref{sec:substrate}, we define
the substrate. In section~\ref{sec:NI}, we study the diffusion on 
non-interacting particles. Hard-core interactions, are introduced in 
section~\ref{sec:HC}, where we analyze how the substrate size and particle
concentration affect the dynamics, and obtain universal forms for the 
time behavior of the MSD of a single tagged
particle and that of the center of mass. Finally, we draw our conclusions
in section~\ref{sec:conclu}.

\section{The substrate}\label{sec:substrate}
The substrate consists of a one-dimensional lattice, with $M$ sites and 
periodical boundary conditions. The particles only jump between 
nearest-neighbor (NN) sites of the lattice. We express every length in units
of the distance $a$ between NN sites; which is equivalent to set $a=1$.
Based on the model introduced in Ref.~\cite{Padilla2009}, we use a discrete 
infinite set of hopping rates, $q_i$ with $i=0,1,2,...$, which depend on two
free parameters $L$ (integer greater than $1$) and $\lambda$ (real positive).

Starting from $q_0$, which fixes the time unit, the other hopping rates are 
defined recursively by
\begin{equation}
  \begin{centering}
    \frac{q_{0}}{q_{i}}= \frac {q_{0}}{q_{i-1}} + (1+\lambda )^{i-1}
    \lambda L^{i},\;\;\;{\rm for}\; i=1,2,3...\;\;\; ,
  \end{centering} \label{relation}
\end{equation}
and are randomly distributed on the lattice with the probability
distribution given by the weights
\begin{equation}
  \begin{centering}
	  f_{i}= \frac{L-1}{L^{i+1}}\;\; ,\;\;{ \rm for}\; i=0,1,2...\;\;\;.
  \end{centering} \label{freq}
\end{equation}
This corresponds to a fractal distribution of hopping rates, which can
be considered a disordered version of the self-similar one introduced in 
Ref.~\cite{Padilla2009}.

In next sections, we address the diffusion properties of
a set particles on this substrate, starting with a uniform distribution.

\section{Non-interacting particles}\label{sec:NI}
We first consider  the case of independent particles.
We focus on the average behavior of a representative tagged particle, 
as expressed by its mean-square displacement $\Delta^2x_{\mbox{\tiny NI}}(t)$ 
at time $t$.

\subsection{The infinite lattice}
It has been shown that, when $M\to\infty$, this substrate is  a 
disordered fractal for length scales much greater than $L$, and,

\begin{equation}
  \begin{centering}
    \Delta^2x_{\mbox{\tiny NI}}(t) = (2 D_0\,  t)^{2 \nu}\;\;\; , 
  \end{centering}
  \label{free_msd}
\end{equation}
for $\Delta^2x_{\mbox{\tiny NI}}\gg L^2$. In equation (\ref{free_msd}),
$D_0$ is a constant, the RW exponent is given by
\begin{equation}
  \nu = \frac{1}{2+\frac{\log(1+\lambda)}{\log L}}\;\;\;{\rm ,}
  \label{exponent}
\end{equation}
and the oscillations that modulate this overall behavior for a deterministic
fractal are washed out by the random shuffling of hopping rates
(for further details, see ~Ref.\cite{Padilla2011}). 

In the rest of this article, to check our analytical results,
we perform numerical Monte Carlo (MC) simulations for
a substrate with  $L=2$ and $\lambda=1$.
The smallest $L$ value makes the substrate a fractal for
the shortest possible distances and times. Also, from equation (\ref{exponent}),
this choice gives $\nu=1/3$, which is sufficiently different from
both $\nu=1/2$ and $\nu=1/4$; which correspond to single-particle and 
single-file diffusion on a homogeneous substrate, respectively.
The data in figures represent averages over at least $10000$ 
disorder realizations.

Before leaving this section, note that if we rewrite 
equation (\ref{free_msd}) as
\begin{equation}
\Delta^2x_{\mbox{\tiny NI}}(t) = 2 D_{\rm eff}(t)t\;\;\; ,
  \label{deff}
\end{equation}
the effective diffusion coefficient $D_{\rm eff}$
satisfies
\begin{equation}
2 D_{\rm eff}(t)=(2 D_0)^{2\nu} t^{2\nu-1}.
  \label{deff_t}
\end{equation}
Or, in terms of the characteristic RW exploration length 
$\ell(t)=\sqrt{\Delta^2x_{\mbox{\tiny NI}}(t)}=(2 D_0\, t)^{\nu}$,

\begin{equation}
 D_{\rm eff}(\ell)=D_0\, \ell^{2-1/\nu},
  \label{deff_l}
\end{equation}
which allows us to interpret $D_{\rm eff}$ as a length-dependent diffusion
 coefficient, and subdiffusion as a result of the power-law  decrease of
the function $D_{\rm eff}(\ell)$.

\subsection{The finite lattice}\label{sub:finite}
For a lattice with a finite number of sites $M$($\gg L$), we expect
that the tagged-particle MSD behaves as in equation (\ref{free_msd}) 
for length scales much shorter than $M$ (though much larger than $L$). 
On the other hand, for length scales much larger 
than $M$, since we work with  periodic boundary conditions, the behavior
is as on a homogeneous substrate. That is, we expect normal diffusion with 
a diffusion coefficient $D_{\rm eff}(\gamma M)$, where $\gamma$ is 
a number of the order of one:

\begin{equation}
	\Delta^2x_{\mbox{\tiny NI}}(t) = 2 D_{\rm eff}(\gamma M)t\;\;\; ,
	  \label{msd_ni_lt}
\end{equation}

We can obtain the crossover time $t_l$ between these two behaviors 
 from the intersections of the right-hand sides of 
equations (\ref{free_msd}) and (\ref{msd_ni_lt}). Using equation 
(\ref{deff_l})), we get

\begin{equation}
	t_l=\frac{(\gamma M)^{1/\nu}}{2D_0},
  \label{t1}
\end{equation}
which is a generalization of the well-known result for a homogeneous
substrate. Indeed, with $\nu=1/2$ in equation (\ref{t1}) we recover
the behavior $t_l\sim M^{2}/{2D_0}$ reported in Ref.~\cite{Beijeren1983}.

\begin{figure}[!ht]
	\begin{center}
    \includegraphics[scale=0.6, clip]{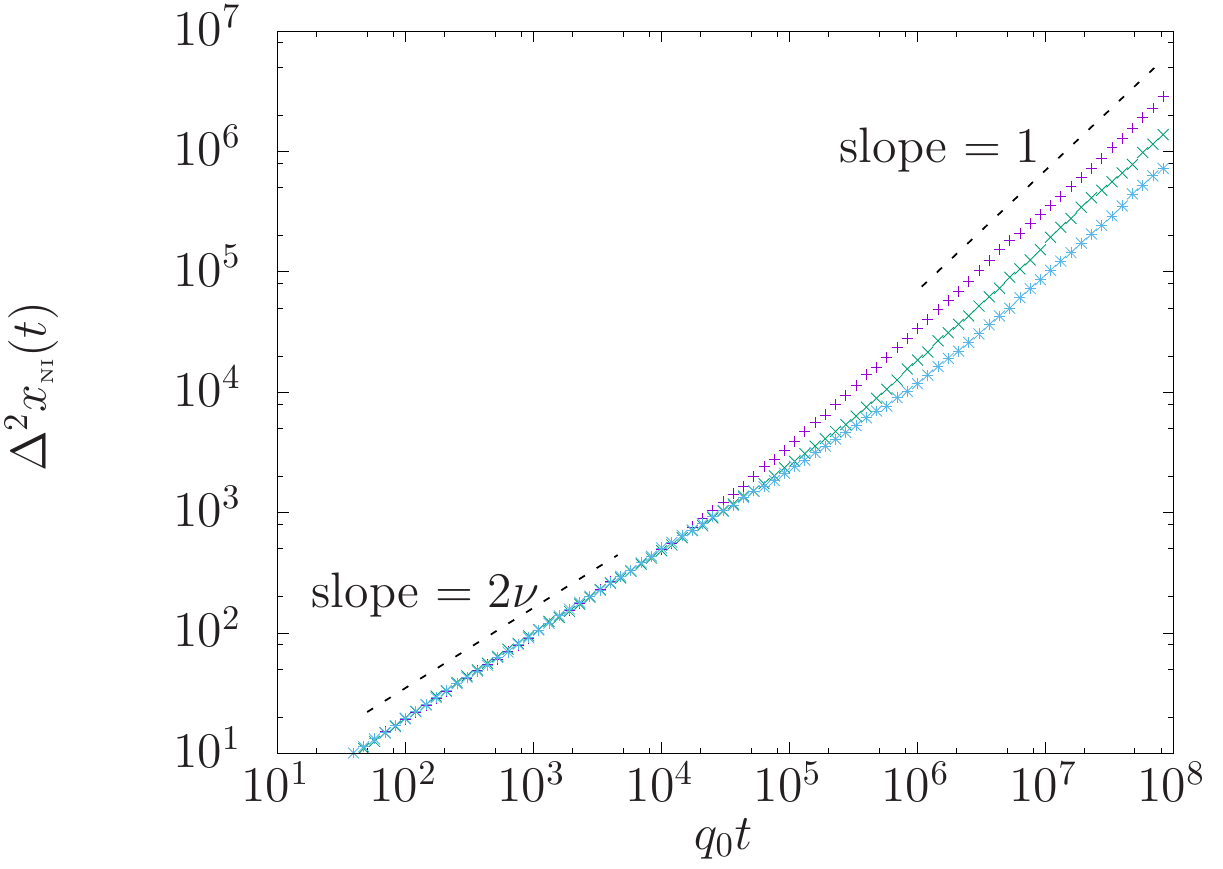}
	 \includegraphics[scale=.6,clip]{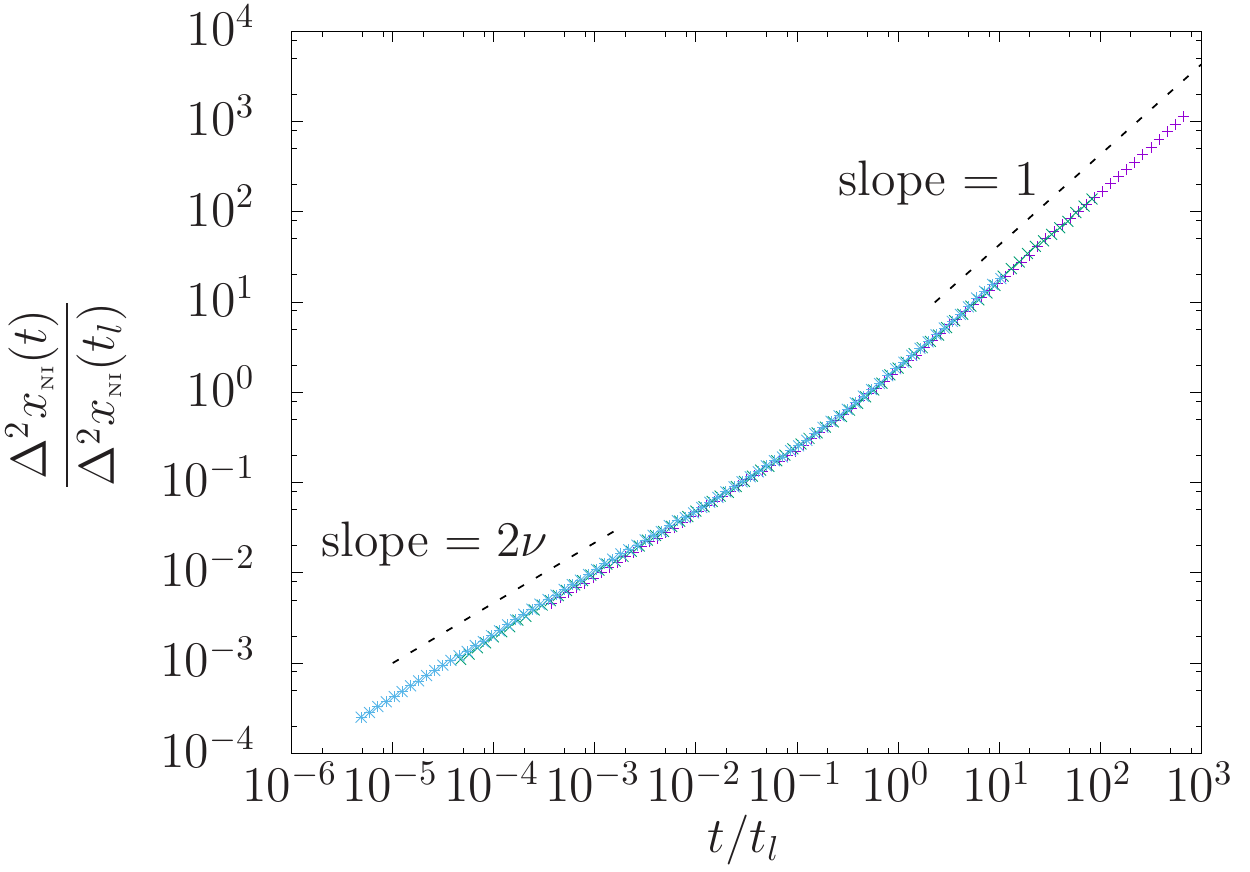}	  
 \end{center}
	\caption{
		(Color online) Dynamics for
		non-interacting particles on a disordered substrate 
		determined by $L=2$ and $\lambda=1$.
		Left: Mean-square displacement of 
	non-interacting particles as a function of time for different lattice 
	sizes; $M=50$ (violet pluses), $M=100$ (green crosses), 
	$M=200$ (light-blue stars). 
	Right: Scaling of the same data with respect to a single variable 
	$t/t_l$. The straight lines indicate the power-law behavior
	in each regime, with $\nu=1/3$.
	
	}
\label{cross1_indep}
\end{figure}

In figure {\ref{cross1_indep}}, we plot the tagged-particle MSD for
various system sizes as a function of time (left panel). In the right
panel of the same figure, we show the same data scaled with respect to
$t/t_l$, using $\gamma=1$. The good collapse on a single curve is 
consistent with the $M$ dependence of  $t_l$ in equation (\ref{t1}).

\section{Hard-core interacting particles}\label{sec:HC}

If we turn on hard-core interactions between particles, the problem lies in the 
familiar case of single-file diffusion.

\subsection{Short times: the effect of particle concentration}\label{sub:short}
Let us start considering an infinite substrate;  $M\to\infty$.
For short enough times, every particle behaves almost as in the non-interacting
case. At these early stages of time evolution, the correlations between 
particle motions are negligible, and we can use a mean-field 
theory~\cite{Suarez2013}. Within this approximation,
the mean-square displacement of a tagged 
particle in a hard-core interacting system $\Delta^2x_{\mbox{\tiny HC}}$
satisfies

\begin{equation}
	\Delta^2x_{\mbox{\tiny HC}}(t)=(1-c)\Delta^2x_{\mbox{\tiny NI}}(t)\;\; ,
  \label{msd_st_0}
\end{equation}
where $c$ is the average concentration of particles. 
Using equation (\ref{free_msd}), we can rewrite relation
between the behaviors in the non-interacting and hard-core-interacting
cases (\ref{msd_st_0})
as
\begin{equation}
	\Delta^2x_{\mbox{\tiny HC}}(t)=(1-c)({2D_0\, t})^{2\nu}\;\;\; .
  \label{msd_st}
\end{equation}
The quantity $(1-c)$ represents the average probability 
of finding an empty site, and then, at short times, the effect 
of hard-core interactions is to slow down diffusion by this factor.

For longer times, after many collisions,  the motions of neighbor particles
become correlated and the so called single-file effect 
makes the diffusion even slower, which is reflected in the diminution of
the RW exponent, from $\nu$ to $\nu/2$. Indeed, by disordering
the hopping rates of a previously deterministic one-dimensional fractal 
substrate, we expect that, for long enough times, the tagged particle MSD
becomes
\begin{equation}
\Delta^2x_{\mbox{\tiny HC}}=\sqrt{\frac{2}{\pi}}\frac{(1-c)}{c} ({2 D_0\, t})^{\nu}\;\;\; 
  \label{msd_mt}
\end{equation}
\cite{Padilla2011, Suarez2014}.

\begin{figure}[!ht]
  \begin{center}
    \includegraphics[scale=.6,clip]{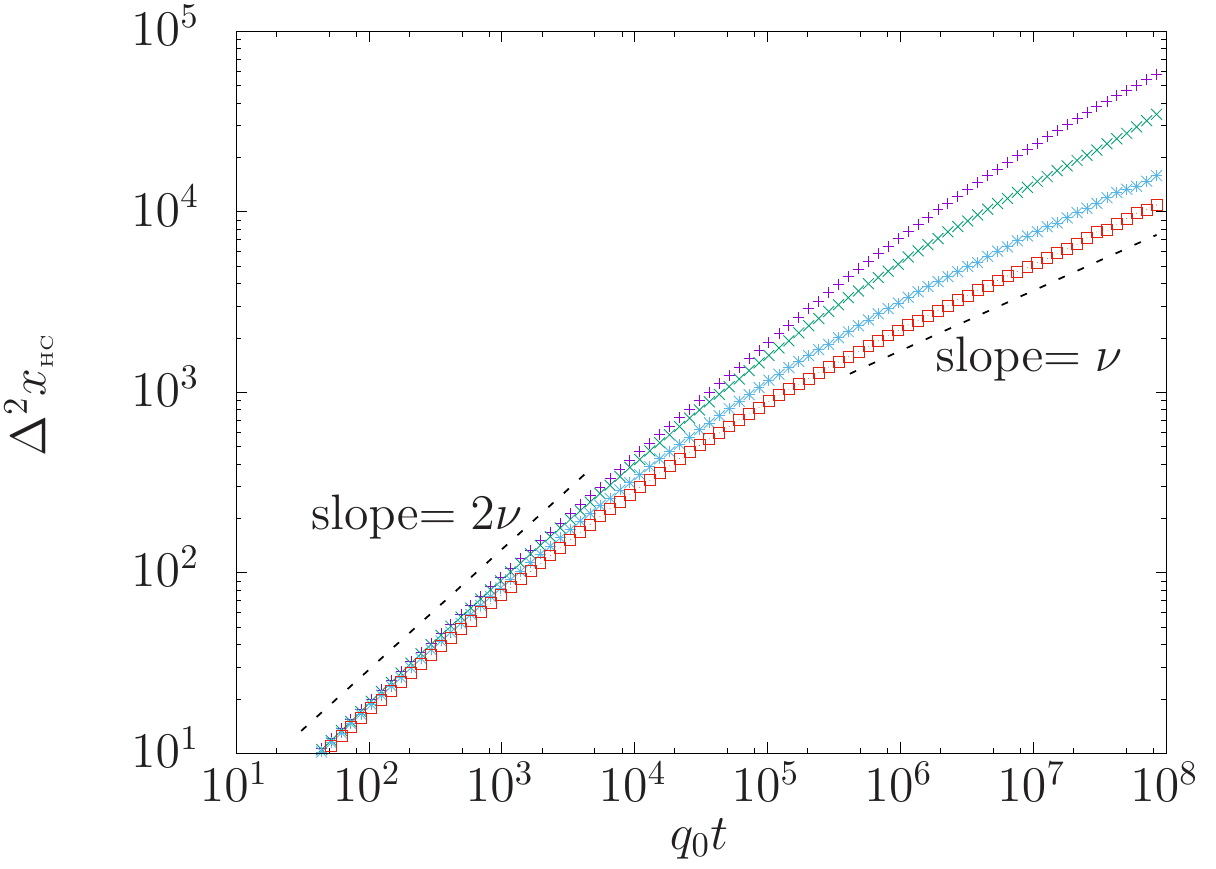}
	\includegraphics[scale=.6,clip]{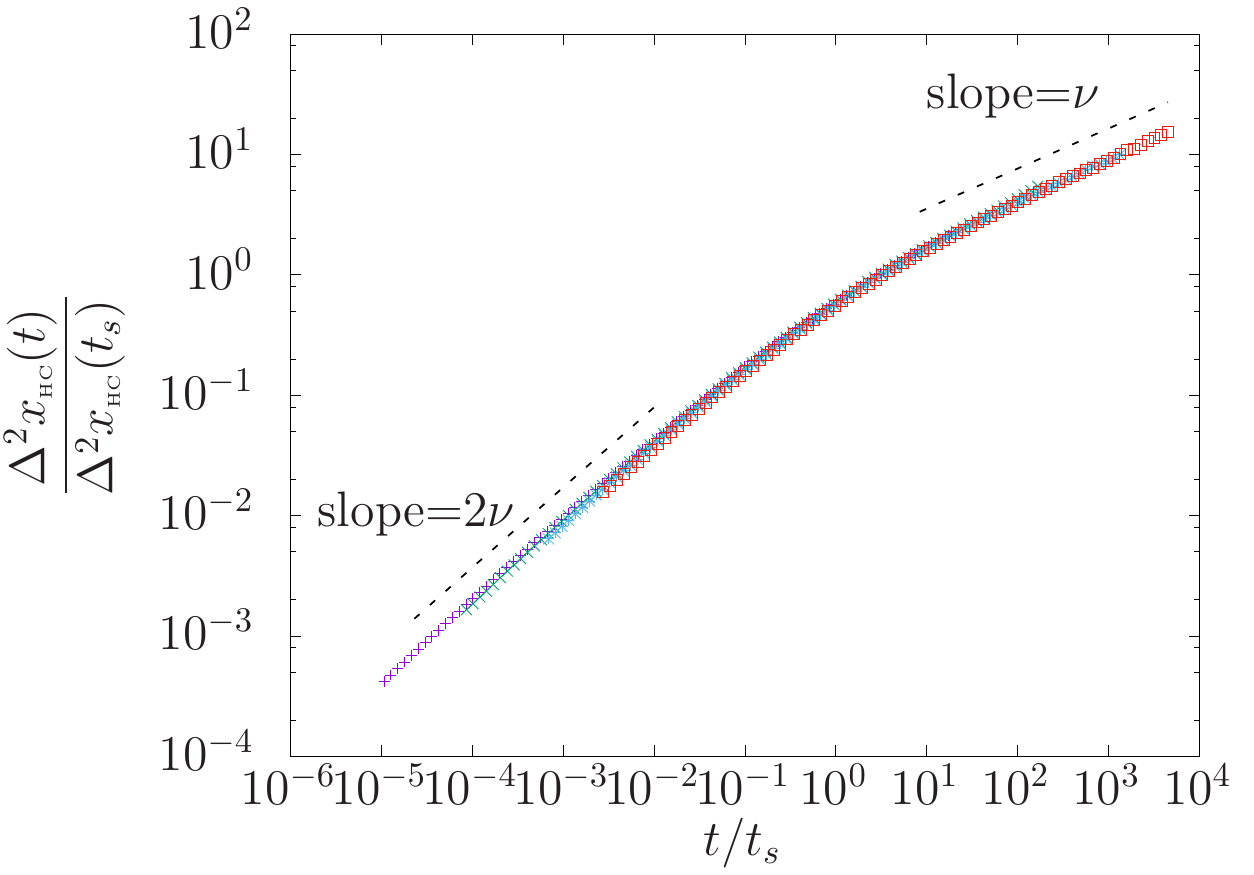}
  \end{center}
	\caption{(Color online)
	Short and intermediate time dynamics for hard-core
	interacting particles.
	Left: Mean-square displacement of a tagged particle for 
	various particle concentrations; $c=5.0\times
	10^{-3}$ (violet pluses), $c=1.0\times 10^{-2}$ (green crosses),
	$c=2.0\times 10^{-2}$ (light-blue stars), $c=3.0\times10^{-2}$ (red 
	squares). Right: Scaling of the same data with respect to a single 
        variable $t/t_s$.}
\label{st_fig}
\end{figure}

In figure~{\ref{st_fig}}-left, we use the results of MC
simulations  to plot the tagged-particle MSD 
as a function of time  for several concentrations. 
The straight lines indicate the predicted short-time 
($\sim t^{2\nu}$) and long-time ($\sim t^\nu$) behaviors, and are drawn to
guide the eyes. We can also see in this figure that the crossover time between
the two dynamical regimes $t_s$ moves to the left as the concentration 
increases. 

We calculate $t_s$ from the intersection of the curves given by 
equations (\ref{msd_st}) and (\ref{msd_mt}). The result is
a crossover time that depends on the concentration of particles:

\begin{equation}
t_s=\frac{1}{2 D_0}\left(\sqrt{\frac{2}{\pi}}\frac{1}{c}\right)^{1/\nu}\;\; .
\label{ts}
\end{equation}
This expression is a generalization of the crossover time $t_s=1/D_0c^2$ that
corresponds to a homogeneous substrate ($\nu=1/2$)~\cite{Suarez2013}.

In the right panel of figure~{\ref{st_fig}}, we plot in a scaled form
the same data shown in left panel,  i.~e., 
$\Delta^2x_{\mbox{\tiny HC}}(t)/\Delta^2x_{\mbox{\tiny HC}}(t_s)$ 
against $t/t_s$. The good collapse on a single curve confirms the 
value of  $t_s$ given by equation (\ref{ts}).

\subsection{Long times: the effect of substrate length}\label{sub:long}

As we discussed in section~\ref{sub:short}, at intermediate times, the MSD
of a tagged particle evolves as $\Delta^2x_{\mbox{\tiny HC}}(t)\sim t^\nu$. 
It is however expected that, for even longer times, the substrate size 
plays a role in the dynamics.  We first investigate  the 
effect of finite $M$ on the time evolution of the center of mass. 
We will see how this analysis helps us to understand the long-time 
behavior of a tagged particle.

Previous studies suggest that, for an initial uniform distribution, 
the introduction o hard-core interactions only affects the diffusion of 
the center of mass by slowing it down by a factor $(1-c)$. 
More precisely, that the relation between the center of mass MSD
for non-interacting and hard-core interacting particles is
\begin{equation}
\Delta^2\bar{x}_{\mbox{\tiny HC}}=(1-c)\Delta^2\bar{x}_{\mbox{\tiny NI}}\;\; .
  \label{msd_cm_1}
\end{equation}
This is the indeed the case of diffusion on an homogeneous 
lattice~\cite{Terranova2005}, 
and on a one-dimensional deterministic fractal~\cite{Suarez2014} when
the oscillatory modulation is averaged out. As the substrate we 
study here can be considered a disordered version of the last, we expect
also in our case, the validity of equation (\ref{msd_cm_1});
which, tacking  into account that for $N$ non-interacting particles 
the relation between the MSD of a tagged particle and of a center of mass is  
$ \Delta^2\bar{x}_{\mbox{\tiny NI}}(t)= \Delta^2 x_{\mbox{\tiny NI}}(t)/N $,
can be rewritten (\ref{msd_cm_1}) as

\begin{equation}
	\Delta^2\bar{x}_{\mbox{\tiny HC}}={(1-c)\over N}\Delta^2{x}_{\mbox{\tiny NI}}\;\; .
	  \label{msd_cm_2}
\end{equation}

\begin{figure}[!ht]
  \begin{center}
  \includegraphics[scale=.8,clip]{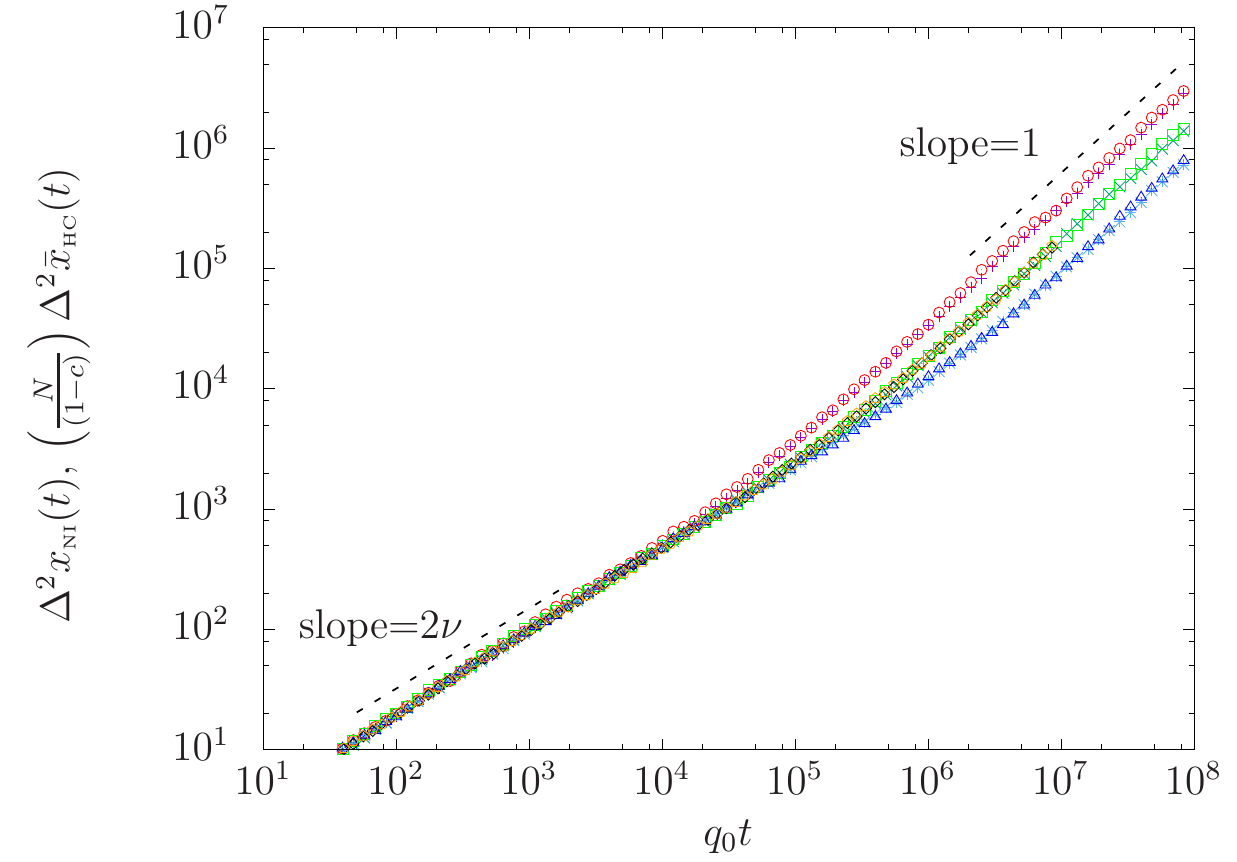}
  \end{center}
  \caption{(Color online) Effects of interactions on
	evolution of the center of mass. 
	With filled symbols, MSD of a tagged particle of 
	non-interacting system: $M=50$ and $c=6.0\times 10^{-1}$ (violet
	pluses), $M=100$ and $c=6.0\times 10^{-1}$ (green crosses), $M=200$ 
	and $c=6.0\times 10^{-1}$ (light-blue stars). With open symbols,
	MSD of the center of mass of hard-core interacting particles, times
	$N/(1-c)$: $M=50$ and $c=6.0\times 10^{-1}$ (red circles), $M=100$ and 
	$c=3.0\times 10^{-1}$ (black diamonds), $M=100$ and 
	$c=3.0\times 10^{-1}$ (light-green squares), $M=100$ and 
	$c=8.0\times 10^{-1}$ (orange pentagons), $M=200$ and 
	$c=6.0\times 10^{-1}$ (blue triangles). 
	}
       \label{cm_effect}
\end{figure}

In figure~\ref{cm_effect}, we show the numerical results that correspond
to the time evolution of $\Delta^2{x}_{\mbox{\tiny NI}}$ (filled symbols) and 
$ {\left( N/( 1-c)\right)}\Delta^2{x}_{\mbox{\tiny HC}}$ 
(open symbols), for several 
concentrations and system sizes (see figure caption). The collapse
of the data for every $M$ shows that  equations (\ref{msd_cm_1}) 
and (\ref{msd_cm_2}) also hold for a disordered one-dimensional fractal.
As a consequence, since for a given system size
and concentration the factor ${\left( ( 1-c)/N)\right)}$ in equation
(\ref{msd_cm_2}) is a constant, the center-of-mass MSD behavior
crosses over from $\sim t^{2\nu}$ to $\sim t$ at the same time $t_l$
as in equation (\ref{t1}), despite  hard-core interactions.

Now, if for not short times, we compare the behavior of a 
tagged particle and the center of mass, we note that 
$\Delta^2\bar{x}_{\mbox{\tiny HC}}(t)$ ($\sim t^{2\nu}$ or $\sim t$) grows 
faster than $\Delta^2x_{\mbox{\tiny HC}}(t)$ ($\sim t^\nu$, 
as described by equation (\ref{msd_mt})). 
This cannot, however, hold forever, because the
MSD of the center of mass should always be smaller or equal than the MSD of 
a single particle. The saturation occurs when all particles follow 
the same behavior. 

\begin{figure}[!ht]
  \begin{center}
     \includegraphics[scale=.6,clip]{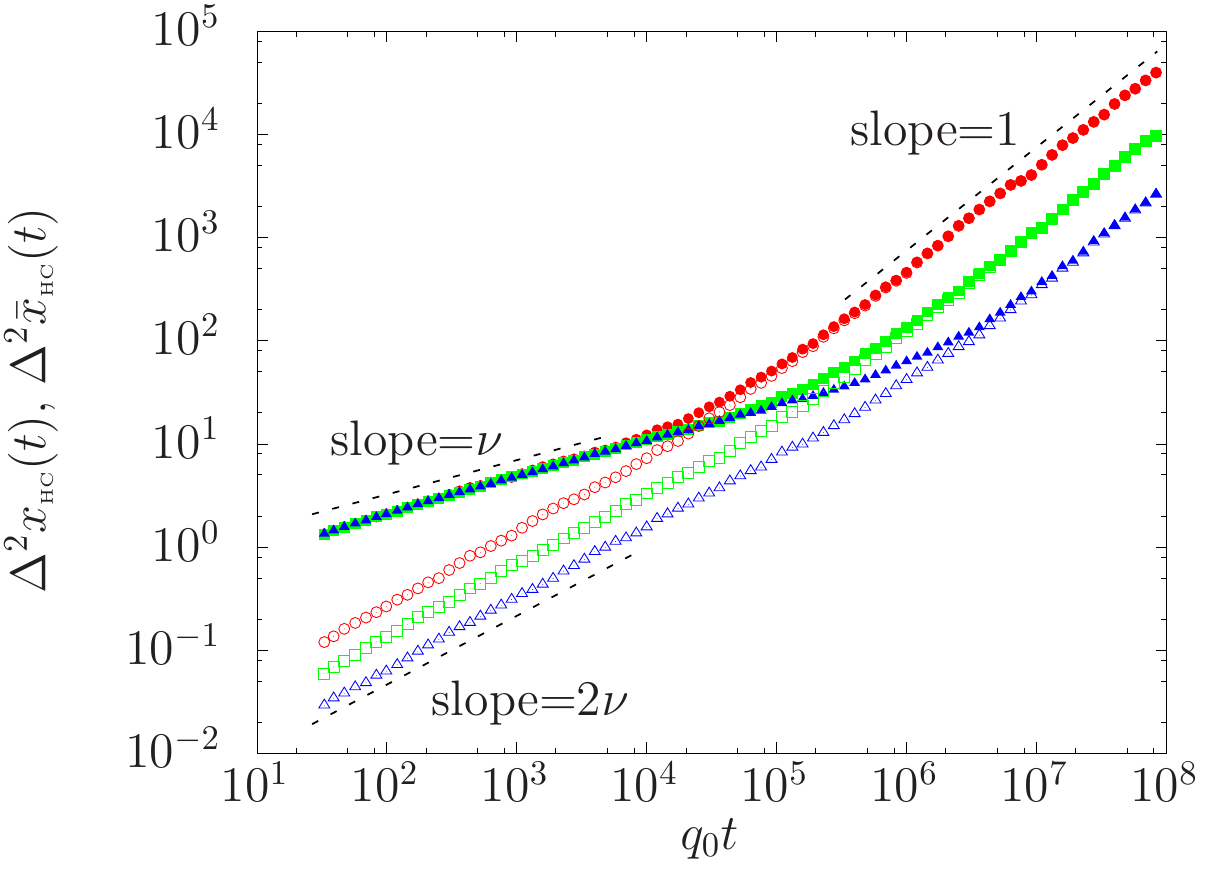}
 \includegraphics[scale=.6,clip]{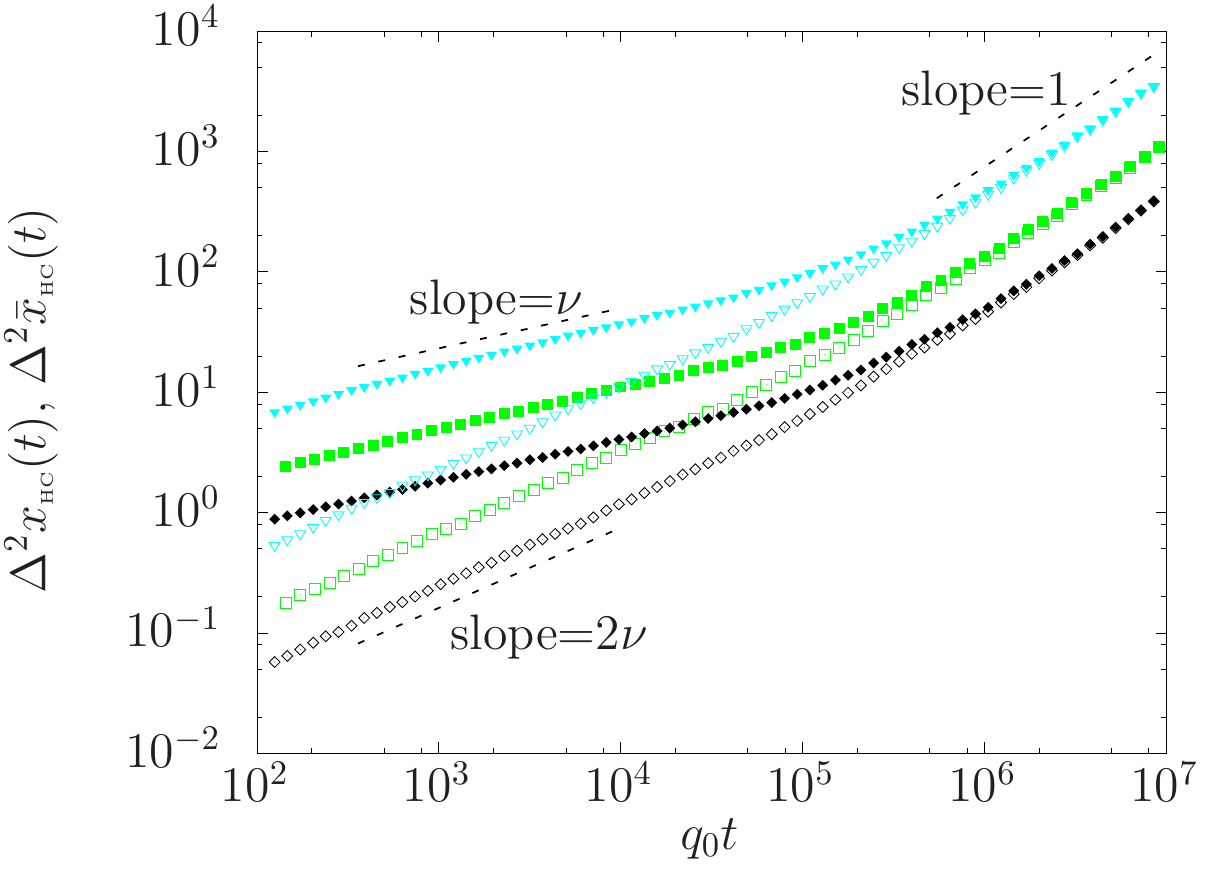}
 \end{center}
\caption{(Color online)
	Intermediate and long time dynamics of a hard-core interacting system.
	MSD for both a tagged particle 
	(filled symbols) and the center of mass (open symbols)
	 Left: Numerical results for a fixed particle concentration
	$c=6.0\times 10^{-1}$, and  various system sizes, $M=50$ (red circles)
	$M=100$ (green squares), $M=200$ (blue upwards triangles).
	Right: Corresponding results for a fixed system size $M=100$ and
	various particle concentrations, $c=3.0\times 10^{-1}$ (cyan downwards 
	triangles), $c=6.0\times 10^{-1}$ (green squares), 
	$c=8.0\times 10^{-1}$ 
	(black diamonds)}
	\label{lt_fig}
\end{figure}

In figure \ref{lt_fig}, we plot  the MSD of both a tagged particle (filled
symbols) and  the center of mass (open symbols). In left panel, we use
a fixed concentration $c=6\times 10^{-1}$ and $M=50, 100, 200$. In
right panel, we use a fixed system size $M=100$ and $c=3\times 10^{-1}, 
6\times 10^{-1}, 8\times 10^{-1}$.
We observe that, the longer the elapsed time, the closer the values 
of $\Delta^2\bar{x}_{\mbox{\tiny HC}}$ and $\Delta^2x_{\mbox{\tiny HC}}$. 
This is an indication of a growing correlation
length, which measures the average size of a box with particles moving cooperatively.
According to this, when, at time $t_l$, this growing length is of the order of system size $M$, 
two important changes occur.
On the one hand, all particles start behaving similarly, which implies 
$\Delta^2\bar{x}_{\mbox{\tiny HC}}(t)\simeq\Delta^2x_{\mbox{\tiny HC}}(t)$.
On the other hand, every particle has already acquired information on the whole
substrate, and starts to diffuse normally as on a periodic chain; 
 $\Delta^2x_{\mbox{\tiny HC}}(t)\sim t$. Thus, we can use $t_l$ also as
 a good estimate for the 
 instant at which the exponent leading the time behavior of the
 tagged-particle MSD crosses over from $\nu$ to $1$. 
  Notice that $t_l$ does not depend on the concentration of particles, 
  in agreement with the plots in figure \ref{lt_fig}-right. 

\begin{figure}[!ht]            
 \begin{center}
 \includegraphics[scale=.8,clip]{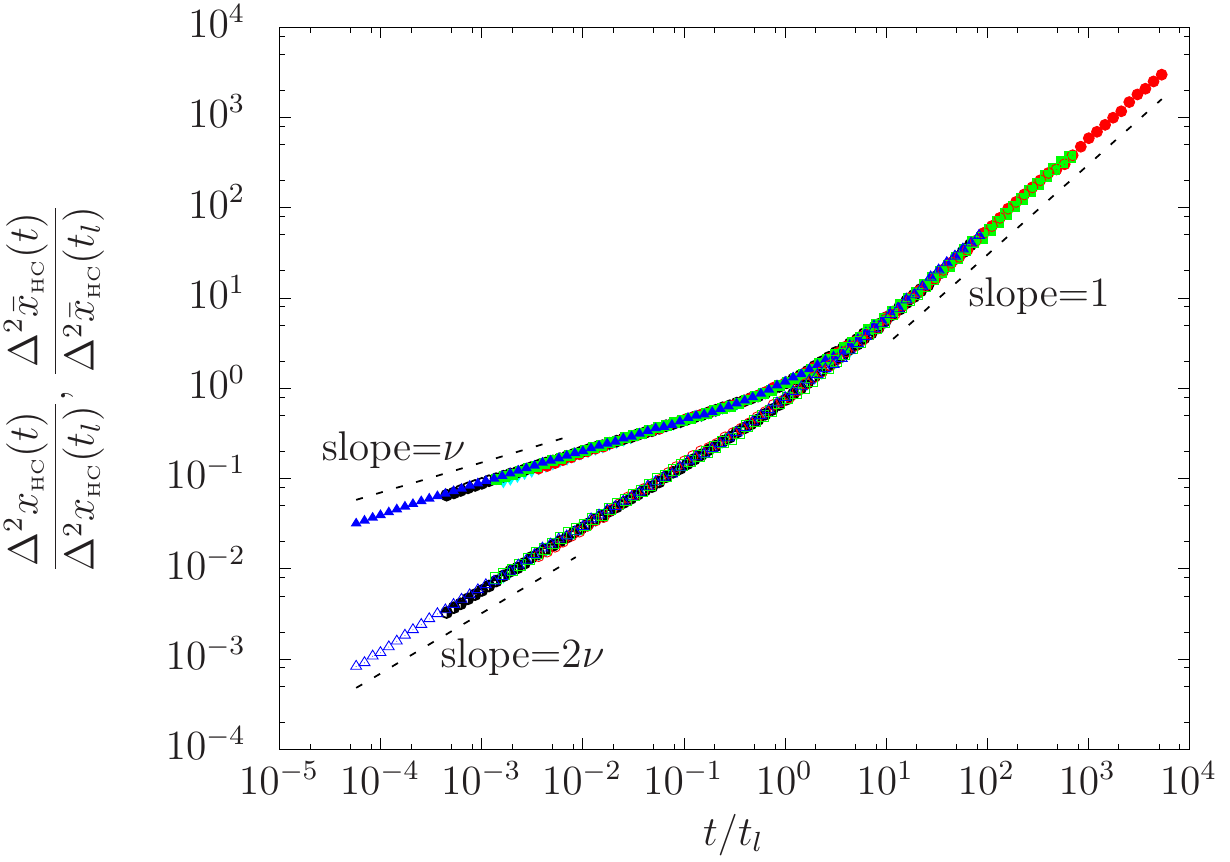}
 \end{center}
	\caption{(Color online) Intermediate-time and long-time 
	collapse of the MSD of both a tagged particle and the center of 
	mass. 
	Scaling of the same data in figure \ref{lt_fig} with respect to a 
	single variable $t/t_l$, with $\gamma=1$. 
	Symbols and colors are the same as in that figure. }
	\label{scal2_fig}
\end{figure}

In figure \ref{scal2_fig}, we show,  using the same data as in 
figure \ref{lt_fig}, the tagged-particle and center-of-mass MDS's
time behaviors, scaled with respect to a single variable $t/t_l$ ($\gamma=1$).
The collapse of the data on a single curve for each observable, confirms that
$t_l$ is a good estimate for the intermediate-time to long-time crossover 
of $\Delta^2x_{\mbox{\tiny HC}}(t)$ as it is for the crossover of 
$\Delta^2\bar{x}_{\mbox{\tiny HC}}(t)$.

\subsection{Universal form for the mean-square displacement}\label{sub:universal}
The above results allow us to state the following universal form for the
MSD of the center of mass:
\begin{equation}
	\Delta^2\bar{x}_{\mbox{\tiny HC}}(t)=\Delta^2\bar{x}_{\mbox{\tiny HC}}
	(t_l)\bar{\cal G}(t/t_l)\;\;\; ,
	\label{univ_cm}
\end{equation}
where $\bar{\cal G}(x)$ is a universal scaling function.

\begin{figure}[!ht]
  \begin{center}
  \includegraphics[scale=.6,clip]{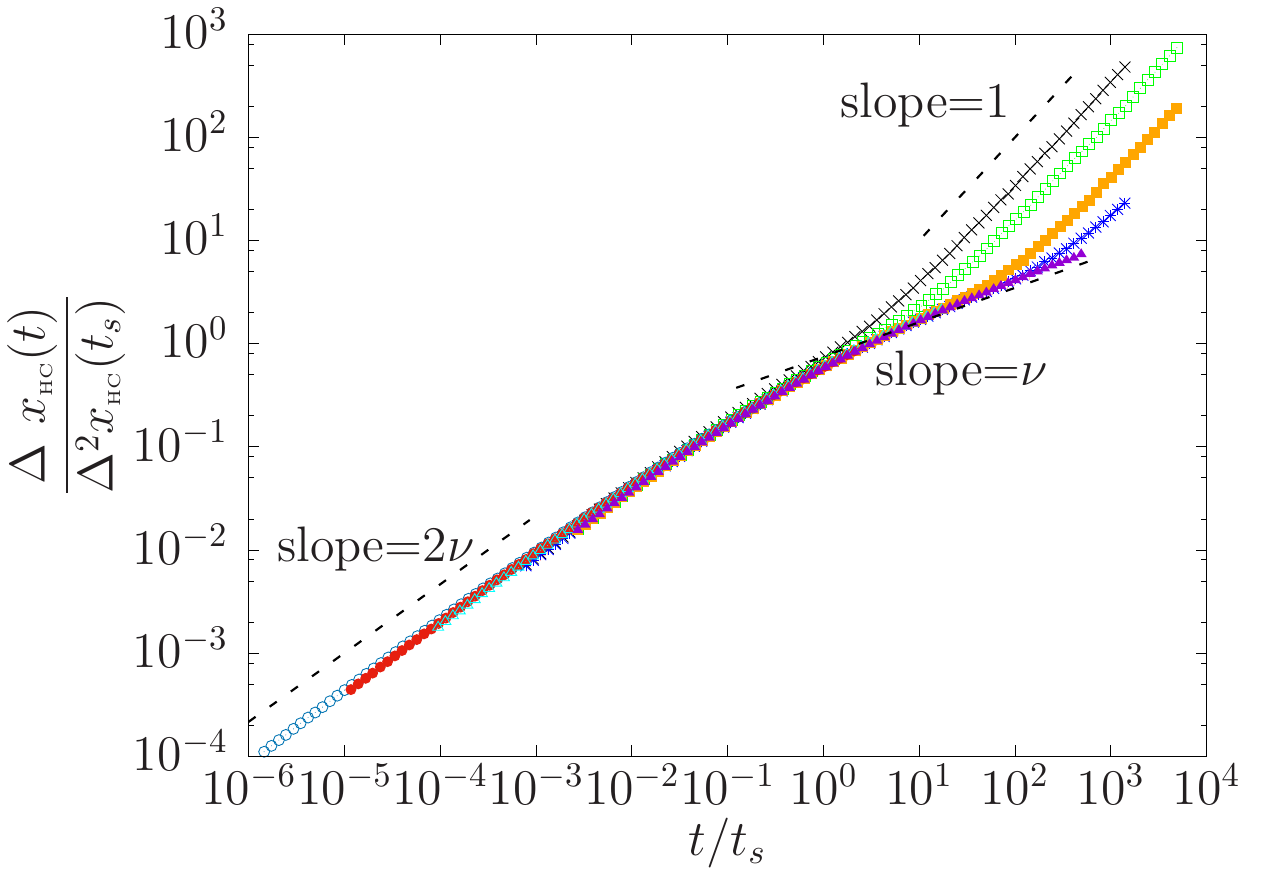}
  \includegraphics[scale=.6,clip]{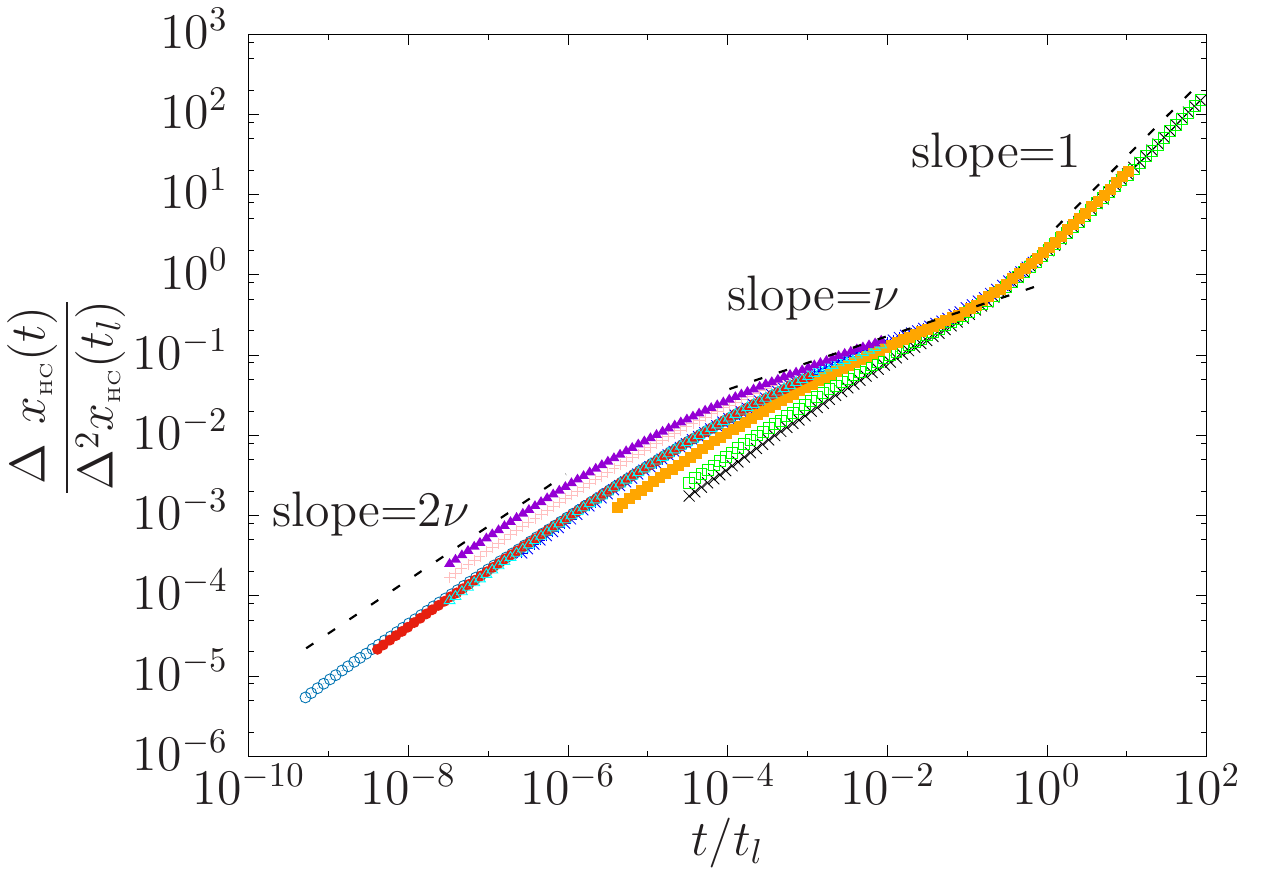}
   \end{center}
	\caption{(Color online) 
	Short, intermediate, and long time MSD of a tagged particle
	in a system with hard-core interactions.
	Left: Short-time and intermediate-time collapse of 
	the results obtained for several particle concentrations and 
	lattice sizes.
	Right: Intermediate-time and long-time collapse of the same results.
	Symbols in both panels:	$c=2.0\times10^{-2}$ and $M=1000$ (pink pluses),
	$c=2.0\times10^{-2}$ and $M=100$ (black crosses), $c=2.0\times10^{-2}$
	 and $M=500$ (blue stars), $c=3.0\times10^{-2}$ and $M=100$ (green 
	 open squares), $c=3.0\times10^{-2}$ and $M=200$ (orange filled 
	 squares), $c=2.5\times10^{-3}$ and $M=4000$ (blue open circles),
	$c=5.0\times10^{-3}$ and $M=2000$ (red filled circles),
	$c=1.0\times10^{-2}$ and $M=1000$ (cyan open triangles),
	$c=3.0\times10^{-2}$ and $M=1000$ (violet filled triangles).
	
	}
 \label{scal_1-y-2_fig}
\end{figure}

The dynamics of a tagged particle is richer, and show two crossovers 
instead of one, at times $t_s$ and $t_l$. As a consequence, it is not
possible to obtain a universal scaling form, valid for every time
scale, by simply scaling with respect to
a single variable. For instance, the scaling of the tagged-particle MSD 
with respect to $t/t_s$  leads to a collapse on a single curve for short 
and intermediate times ($t\ll t_l$) which disappears at longer times 
(figure \ref{scal_1-y-2_fig}-left). Conversely, the scaling of the same
observable with respect to $t/t_l$ produces a good collapse of the data
for long and  intermediate times ($t\gg t_s$), which disappears at shorter
times (figure \ref{scal_1-y-2_fig}-right).

\begin{figure}[!ht]
  \begin{center}
   \includegraphics[scale=.8,clip]{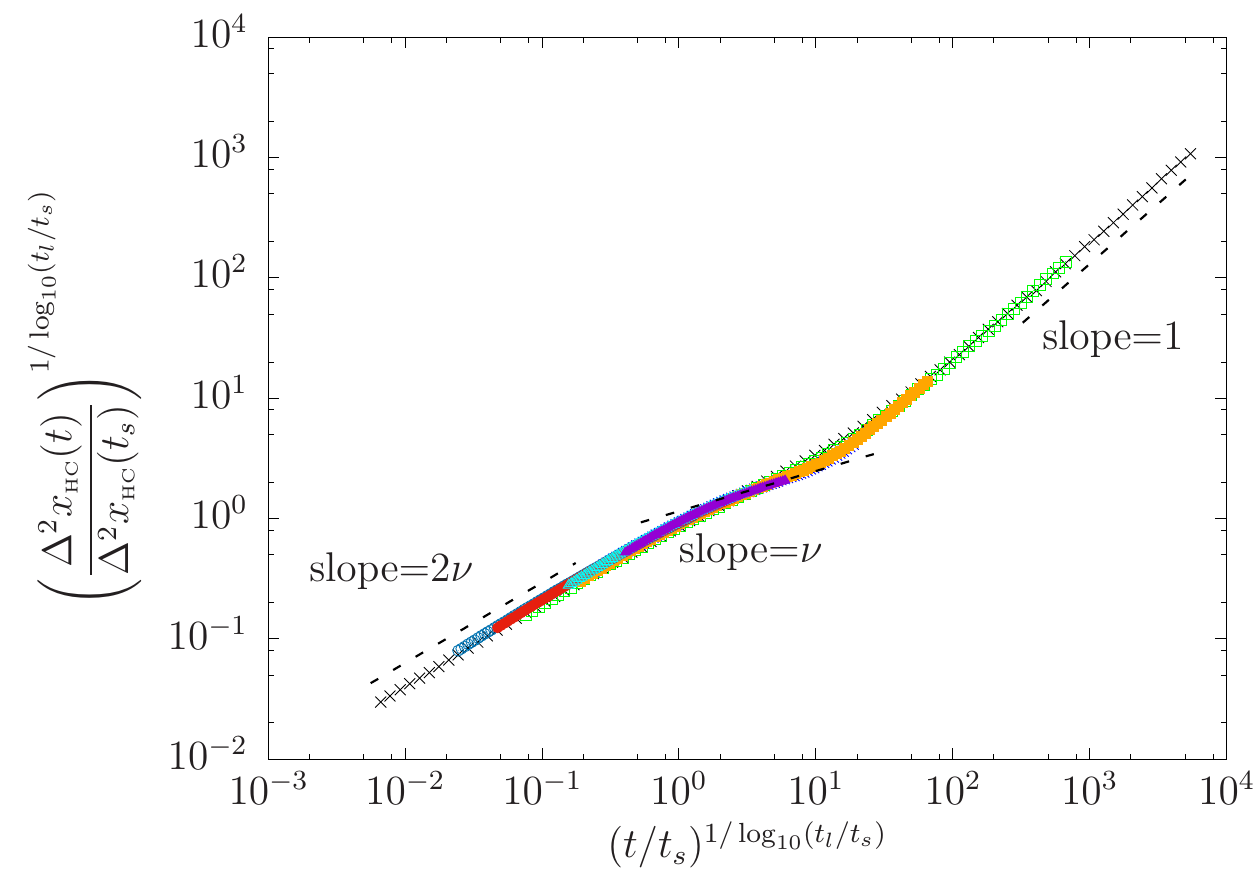}
 \end{center}
	\caption{Scaling form of the MSD of a tagged particle for the same 
	data in figure~\ref{scal_1-y-2_fig}: In this plot, we use
$t_l=\frac{M^{1/\nu}}{2D_0}$ and 
$t_s=\frac{1}{2 D_0}\left(\sqrt{\frac{2}{\pi}}\frac{1}{c}\right)^{1/\nu}$.
	The dashed lines stand for the power-laws leading the  short,
	intermediate, and long time behaviors (from left to right).
	See the main text for further details.}

 \label{scal_total_fig}
\end{figure}

For other systems, the full collapse of functions with two different crossovers 
has be studied in the past (see, for example, 
\cite{Chou2009,Centres2010, Mazzitello2015}). In the case that concerns us,
the universal form of the tagged-particle MSD
is achieved in two steps; a translation followed by an isotropic 
change of scale. 
First, every curve in $\log-\log$
scale is rigidly translated, to move the first 
(short to intermediate times) crossover point to the origin, as
we do in figure~\ref{scal_1-y-2_fig}-left, i.~e., 
by plotting $\left(\displaystyle\frac{\Delta^2x_{\mbox{\tiny HC}}(t)}{\Delta^2x_{\mbox{\tiny HC}}(t_s)}\right)$ as a function of $t/t_s$. 
	Then, both axis are rescaled by the same factor 
	$1/\log\left(\frac{t_l}{t_s}\right)$. Both operations conserve the 
        slopes in the three regimes.	
	The plot that results after the whole transformation, for the data 
	in figure~\ref{scal_1-y-2_fig}, is shown in figure~\ref{scal_total_fig}.
	The very good collapse on a single curve is apparent, and gives 
	support to the idea of universality, according to which, the MSD of a 
	tagged particle satisfies

\begin{equation}
 \log\left(\displaystyle\frac{\Delta^2x_{\mbox{\tiny HC}}(t)}
	{\Delta^2x_{\mbox{\tiny HC}}(t_s)}
	\right)=
	{\displaystyle\log\left(\frac{t_l}{t_s}\right)}
	{\cal F}\left[\displaystyle{\log\left(\frac{t}{t_s}\right)}\over
	{\displaystyle\log\left(\frac{t_l}{t_s}\right)}
	\right]\;{\mbox ,}
	\label{univ_f}
\end{equation}	
where ${\cal F}(x)$ is an universal scaling function. This can also
be expressed as
\begin{equation}
	\left({\Delta^2x_{\mbox{\tiny HC}}(t)}\right)^\mu=
	    \left({\Delta^2x_{\mbox{\tiny HC}}(t_s)}\right)^\mu
	 {\cal G}\left[\displaystyle{\left(t\over t_s\right)^\mu}\right]
	 \;{\mbox ,}
							         \label{univ_g}
\end{equation}
where ${\cal G}(x)$ is another scaling function, and $\mu=1/\log\left(\frac{t_l}{t_s}\right)$

\section{Conclusion}\label{sec:conclu}

In this work, we study single-file diffusion on a disordered fractal. 
More specifically, we focus on the MSD time behavior for the center of mass and 
 a tagged particle. We use one-dimensional substrates of finite length, 
with periodic boundary conditions, where the dynamics are richer than for
infinite substrates.

We propose a general relation between the center-of-mass MSD  
for a set of hard-core interacting particles, and the MSD of a 
single particle on the substrate; which we then verify numerically
for all times. We can interpret that the factor $(1-c)/N$ between
these two behaviors, comes from the fact 
that the system has $N$ particles, and that the hopping rates 
are reduced by a factor $(1-c)$ due to single occupancy. 
We expect that the same relation will be valid for other fractal substrates. 

We show that the behavior of the center-of-mass MSD 
presents two regimes; subdiffusion ($\sim t^{2\nu}$) at short
times, and normal diffusion ($\sim t$) at long times.
Where, $\nu$ is the RW exponent of a single particle on the same fractal
structure of infinite length.
Note that for an homogeneous substrate ($\nu=1/2$),
short-time and long-time exponents coincide, and the center of
mass diffuses always normally, as known.

The results of the center-of-mass MSD for different lattice 
lengths and concentrations collapse on a single scaling form, 
where the crossover time, which does not depend 
on concentration, is the same as for non-interacting particles.

The MSD of a tagged particle presents three time regimes. In the first 
two, the dynamics are subdiffusive; with RW exponents $2\nu$ and $\nu$, 
respectively. 
At short times no correlation between particles exists,
and the tagged-particle MSD  behavior is almost 
the same as for a single particle; only modified by
the mean-field factor $(1-c)$. 
At intermediate times, the behavior is 
more complex, and because of correlations, the RW exponent is 
reduced from $2\nu$ to $\nu$.   
We find an analytical expression for the corresponding crossover
time, which depends on particle concentration but not on substrate
length. This dependence is confirmed numerically. 
For long enough times, the  tagged 
particle behaves as the center of mass, and reaches the normal 
diffusive regime. We predict that the crossover time between the
second and third regimes has the same substrate-length dependence
as the crossover time for the center-of-mass MSD, and that it does
not depend on particle concentration. The outcomes of simulations confirm
this prediction.
The knowledge of the two crossover times allows us
to formulate the universal scaling form  for the tagged-particle MSD 
on a disordered fractal, which generalizes the one for 
an homogeneous substrate.

Let us finally remark that, in the infinite system size limit, 
the MSD of a tagged particle presents 
only the first two time regimes,
and the dynamics of the center of mass are always subdiffusive, with RW 
exponent $2\nu$.

\section*{Acknowledgments}
We acknowledge useful discussions with Gonzalo Su\'arez.
This research was supported in part by the Consejo Nacional de Investigaciones
Cient\'{\i}ficas y  T\'ecnicas (CONICET), and the
Universidad Nacional de Mar del Plata. JLI is grateful for the financial
support and hospitality of the Abdus Salam International Centre for Theoretical
Physics (ICTP), where part of this work was developed.

\section*{References}

\begin{thebibliography}{10}

\bibitem{havlin1987}
Shlomo Havlin and Daniel Ben-Avraham.
\newblock Diffusion in disordered media.
\newblock {\em Advances in Physics}, 36(6):695--798, 1987.

\bibitem{rammal1983}
{Rammal, R.} and {Toulouse, G.}
\newblock Random walks on fractal structures and percolation clusters.
\newblock {\em J. Physique Lett.}, 44(1):13--22, 1983.

\bibitem{alexander1982}
{Alexander, S.} and {Orbach, R.}
\newblock Density of states on fractals : `` fractons ''.
\newblock {\em J. Physique Lett.}, 43(17):625--631, 1982.

\bibitem{bouchaud1990}
Jean-Philippe Bouchaud and Antoine Georges.
\newblock Anomalous diffusion in disordered media: Statistical mechanisms,
  models and physical applications.
\newblock {\em Physics Reports}, 195(4?5):127 -- 293, 1990.

\bibitem{ben2000diffusion}
D.~Ben-Avraham and S.~Havlin.
\newblock {\em Diffusion and Reactions in Fractals and Disordered Systems}.
\newblock Cambridge University Press, 2000.

\bibitem{Grabner1997}
Peter~J. Grabner and Wolfgang Woess.
\newblock Functional iterations and periodic oscillations for simple random
  walk on the Sierpinski graph.
\newblock {\em Stochastic Processes and their Applications}, 69(1):127 -- 138,
  1997.

\bibitem{acedo2000}
L.~Acedo and S.~B. Yuste.
\newblock Territory covered by n random walkers on fractal media: The
  Sierpinski gasket and the percolation aggregate.
\newblock {\em Phys. Rev. E}, 63:011105, Dec 2000.

\bibitem{bab2008EPL}
M.~A. Bab, G.~Fabricius, and E.~V. Albano.
\newblock On the occurrence of oscillatory modulations in the power law
  behavior of dynamic and kinetic processes in fractals.
\newblock {\em EPL (Europhysics Letters)}, 81(1):10003, 2008.

\bibitem{bab2008JCP}
M.~A. Bab, G.~Fabricius, and Ezequiel~V. Albano.
\newblock Revisiting random walks in fractal media: On the occurrence of time
  discrete scale invariance.
\newblock {\em The Journal of Chemical Physics}, 128(4):--, 2008.

\bibitem{maltz2008}
Alberto~L. Maltz, Gabriel Fabricius, Marisa~A. Bab, and Ezequiel~V. Albano.
\newblock Random walks in fractal media: a theoretical evaluation of the
  periodicity of the oscillations in dynamic observables.
\newblock {\em Journal of Physics A: Mathematical and Theoretical},
  41(49):495004, 2008.

\bibitem{weber2010}
Sebastian Weber, Joseph Klafter, and Alexander Blumen.
\newblock Random walks on Sierpinski gaskets of different dimensions.
\newblock {\em Phys. Rev. E}, 82:051129, Nov 2010.

\bibitem{Padilla2009}
L.~Padilla, H.~O. M\'{a}rtin, and J.~L. Iguain.
\newblock {Log-periodic modulation in one-dimensional random walks}.
\newblock {\em EPL (Europhysics Letters)}, 85(2):20008, January 2009.

\bibitem{lorena2010}
L.~Padilla, H.~O. M\'artin, and J.~L. Iguain.
\newblock Log-periodic oscillations for diffusion on self-similar finitely
  ramified structures.
\newblock {\em Phys. Rev. E}, 82:011124, Jul 2010.

\bibitem{Padilla2011}
L.~Padilla, H.~M{\'{a}}rtin, and J.~Iguain.
\newblock {Anomalous diffusion with log-periodic modulation in a selected time
  interval}.
\newblock {\em Physical Review E}, 83(2):2--5, Feb 2011.

\bibitem{havlin2000}
Daniel Ben Avraham and Shlomo Havlin.
\newblock {\em Diffusion and Reactions in Fractals and Disordered Systems}.
\newblock Cambridge University Press, 2000.

\bibitem{bernhard2004}
Bernhard Kr\"on and Elmar Teufl.
\newblock Asymptotics of the transition probabilities of the simple random walk
  on self-similar graphs.
\newblock {\em Trans. Amer. Math. Soc.}, 356:393--414, 2004.

\bibitem{Padilla2012}
L.~Padilla, H.~O. M{\'{a}}rtin, and J.~L. Iguain.
\newblock {Anisotropic anomalous diffusion modulated by log-periodic
  oscillations}.
\newblock {\em Physical Review E}, 86(1):011106, Jul 2012.

\bibitem{refId0}
{Frechero, M. A.}, {Padilla, L.}, {M\'artin, H. O.}, and {Iguain, J. L.}
\newblock Intermediate-range structure in ion-conducting tellurite glasses.
\newblock {\em EPL}, 103(3):36002, 2013.


\bibitem{Harris1965}
T.~E. Harris.
\newblock {Diffusion with ``collisions'' between particles}.
\newblock {\em Journal of Applied Probabillity}, 2(2):323--338, 1965.

\bibitem{Percus1974}
Jerome~K. Percus.
\newblock Anomalous self-diffusion for one-dimensional hard cores.
\newblock {\em Physical Review A}, 9(1):1--3, 1974.

\bibitem{Richards1977}
Peter~M. Richards.
\newblock Theory of one-dimensional hopping conductivity and diffusion.
\newblock {\em Phys. Rev. B}, 16:1393--1409, Aug 1977.

\bibitem{Beijeren1983}
Henk van Beijeren, K.~W. Kehr, and R.~Kutner.
\newblock Diffusion in concentrated lattice gases. iii. tracer diffusion on a
  one-dimensional lattice.
  \newblock {\em Phys. Rev. B}, 28:5711--5723, Nov 1983.

\bibitem{Lizana2009}
 L.~Lizana and T.~Ambj\"ornsson.
 \newblock Diffusion of finite-sized hard-core interacting particles in a
    one-dimensional box: Tagged particle dynamics.
 \newblock {\em Phys. Rev. E}, 80:051103, Nov 2009.

\bibitem{Centres2010}
 P.~M. Centres and S.~Bustingorry.
\newblock Effective edwards-wilkinson equation for single-file diffusion.
\newblock {\em Phys. Rev. E}, 81:061101, Jun 2010.

\bibitem{Suarez2013}
Gonzalo Su{\'{a}}rez, Miguel Hoyuelos, and H{\'{e}}ctor~O. M{\'{a}}rtin.
\newblock {Evolution equation for tagged-particle density and correlations in
single-file diffusion}.
\newblock {\em Physical Review E - Statistical, Nonlinear, and Soft Matter
Physics}, 88(2):1--7, 2013.

\bibitem{Suarez2014}
G~P Su{\'{a}}rez, H~O M{\'{a}}rtin, and J~L Iguain.
\newblock {Single-file diffusion on self-similar substrates}.
\newblock {\em Journal of Statistical Mechanics: Theory and Experiment},
 2014(7):P07010, jul 2014.

 \bibitem{Terranova2005}
 G.~Terranova, H.~M\'{a}rtin, and C.~Aldao.
 \newblock {Exact diffusion coefficient for a chain of beads in one dimension
 using the Einstein relation}.
\newblock {\em Physical Review E}, 72(6):061108, December 2005.

\bibitem{Chou2009}
Yen-Liang Chou and Michel Pleimling.
\newblock {Parameter-free scaling relation for nonequilibrium growth
 processes}.
\newblock {\em Physical Review E}, 79(5):051605, may 2009.

 \bibitem{Mazzitello2015}
K~I Mazzitello, L~M Delgado, and J~L Iguain.
\newblock {Low-coverage heteroepitaxial growth with interfacial mixing}.
\newblock {\em Journal of Statistical Mechanics: Theory and Experiment},
2015(1):P01015, jan 2015.




\end{thebibliography}
%


\end{document}